\documentclass[12pt,showpacs,superscripta ddress,preprintnumbers,amsmath,amssymb]{revtex4}
\usepackage{epsfig} 
\usepackage{amsmath} 
\usepackage{graphicx} 
\usepackage{subfigure}
\large

\newcommand{\ba}{\begin{array}}
\newcommand{\ea}{\end{array}}
\newcommand{\bea}{\begin{eqnarray}}
\newcommand{\eea}{\end{eqnarray}}
\renewcommand{\l}{\left}
\renewcommand{\r}{\right}

\newcommand{\Ga}{\Gamma}
\newcommand{\De}{\Delta}

\def\ket#1{\left| #1\right\rangle}

\begin{document} 

\title{Decoherence free $B_{d}$ and $B_{s}$ meson systems }

\author{Ashutosh Kumar Alok}
\email{akalok@iitj.ac.in}
\affiliation{Indian Institute of Technology Jodhpur, Jodhpur 342011, India}

\author{Subhashish Banerjee}
\email{subhashish@iitj.ac.in}
\affiliation{Indian Institute of Technology Jodhpur, Jodhpur 342011, India}

\date{\today} 
\preprint{}

\begin{abstract}
We study the impact of decoherence on $B$ meson systems with specific emphasis on $B_s$. For consistency we also study the $B_d$ mesons based on 
the most recent data.  We find that the $B_d$ mesons are $34\sigma$ away from total decoherence, while the  $B_s$ mesons are seen to be upto $31\sigma$ away from total decoherence.
Thus, our results prove, with experimental verity, that neutral meson systems are free from decoherence effects. Therefore, this provides a very
useful laboratory for testing the foundations of quantum mechanics.

\end{abstract}

\pacs{14.40.Nd, 13.25.Hw, 3.65.Yz,  03.65.Ta} 

\maketitle 

\newpage

\section{Introduction}
Any system that evolves is affected by its surroundings which could be considered to be its environment. 
When the effect of the environment is taken into account for the dynamics of the system of interest, 
one is lead naturally to phenomenon such as decoherence.
This is an open system way of treating these issues, making the evolution of the quantum system of interest to be
non unitary in general. Such ideas have been fruitfully used in studies in
quantum optics \cite{louisell, bp}, quantum information \cite{nc} and condensed matter physics \cite{path}. Hence, it is a natural question to pose 
whether the $K$ and $B$ mesons produced at the $K$ and $B$ factories, and which
serve as a rich laboratory for probing physics of and beyond the standard model, are affected by such 
open systems effects such as decoherence or not. 

The foundations of quantum mechanics are usually studied in optical or electronic systems. Here the detection efficiency is much lower 
than that of the corresponding detectors at the high energy frontier experiments such as the Large Hadron Collider (LHC). 
Therefore, it will be interesting to test the foundations of quantum mechanics in unstable 
massive systems at high energies over a macroscopic scale ($\sim 10^{-3}$ cm), as provided by, for e.g., the $B$ factories. Thus these 
systems will provide an alternative platform for testing foundations of quantum mechanics.

In this context there have been many attempts to connect foundational aspects such as decoherence 
and quantum coherence over macroscopic distances in these meson systems 
\cite{Datta:1986ut,Bertlmann:1996at,Dass:1997dd,Bertlmann:1997ei,Bertlmann:1999np,AmelinoCamelia:2010me,Apostolakis:1997td,Ambrosino:2006vr,Go:2007ww}.
Here our motivation is to study the impact of decoherence on $B$ meson systems with specific emphasis on $B_s$ mesons where a lot of accurate data is 
recently coming from LHC. For consistency we also study the $B_d$ mesons based on the most recent data. 
We find that these systems are {\it remarkably coherent}, i.e., free from the effect of decoherence. This makes them pristine objects for the study of 
foundations of quantum mechanics.
We also find this view to be corroborated in the $K$ meson systems \cite{Bertlmann:1999np,AmelinoCamelia:2010me,Apostolakis:1997td,Ambrosino:2006vr}. 
This thus provides a remarkable scenario where unitary quantum evolution is supported in real time dynamics.

The paper is organized as follows: In Sec. \ref{sec:def} we discuss the methodology for determining the decoherence in neutral $B$ meson systems. In Sec. \ref{sec:results}, we present the results obtained for the decoherence parameter in $B_d$  and $B_s$ mesons. Finally in Sec. \ref{sec:concl}, we present our conclusions.

\section{The quantum mechanics of neutral $B$ mesons}
\label{sec:def}

Here we study the quantum mechanics of neutral self-conjugate pairs of B mesons.
There are two such systems: $B_d(\overline{B}_d)$ and $B_s(\overline{B}_s)$ mesons. 
In these B mesons there are two flavor eigenstates, which have definite quark content and are
useful for understanding particle production and decay processes. 
Also, there are mass eigenstates pertaining to states with definite mass and lifetime.

In these systems, the mass eigenstates $B_{L}$ and $B_{H}$ ($L$ and $H$ indicate the light
and heavy states, respectively) are admixtures of the flavor
eigenstates $B_q$ and $\overline{B}_q$ ($q=d,\,s$):
\bea 
\ket{B_L} &=& p \ket{B_q} + q \ket{\overline{B}_q} ~, \nonumber\\
\ket{B_H} &=& p \ket{B_q} - q \ket{\overline{B}_q} ~, 
\eea 
with $|p|^2 + |q|^2 =1$.  As a result, the initial flavor eigenstates
oscillate into one another according to the evolution equation
\bea 
i \frac{d}{dt} \l( \ba{c} \ket{B_q(t)} \\ \ket{\overline{B}_q(t)} \ea \r) =
\l(M^q -i \frac{\Ga^q}{2} \r) \l( \ba{c} \ket{B_q(t)}
\\ \ket{\overline{B}_q(t)} \ea \r) ~,
\eea 
where $M=M^\dagger$ and $\Ga=\Ga^\dagger$ correspond respectively to
the dispersive and absorptive parts of the mass matrix. The
off-diagonal elements, $M^q_{12}=M_{21}^{q*}$ and
$\Ga^q_{12}=\Ga_{21}^{q*}$, are generated by $B_q$-$\overline{B}_q$
mixing. We define
\bea 
\Ga_q \equiv \frac{\Ga_H
  + \Ga_L}{2}, \quad \De M_q \equiv   M_H - M_L, \quad \De \Ga_q \equiv   \Ga_L -
\Ga_H ~.
\eea 
Thus the mass difference $\De M_q $ is positive by definition. The ratio $q/p$ is given by
\begin{equation}
\frac{q}{p} = \frac{\De M_q -\frac{i}{2}\De \Ga_q }{2(M^q_{12}-\frac{i}{2}\Ga^q_{12})} = 
\frac{2(M_{12}^{q*}-\frac{i}{2}\Ga_{12}^{q*})}{\De M_q -\frac{i}{2}\De \Ga_q}\,.
\end{equation}

The wavefunction of a $B_q \overline{B}_q$ pair is in the entangled state
\begin{equation}
\ket{\Psi}=\frac{1}{\sqrt 2} \l(\ket{B_q} \otimes \ket{\overline{B}_q}-\ket{\overline{B}_q} \otimes \ket{B_q} \r)\,.
\label{bb1}
\end{equation}
The time evolution of Eq. (\ref{bb1}) can be shown to be
\bea
\ket{B_q(t)} &=& g_{+}(t)\ket{B_q} + \frac{q}{p}\, g_{-}(t)\ket{\overline{B}_q} ~, \nonumber\\
\ket{\overline{B}_q(t)} &=& \frac{p}{q}\, g_{-}(t)\ket{B_q} +  g_{+}(t)\ket{\overline{B}_q} ~,
\eea
where
\bea
g_{+}(t) &=&  e^{-iM_qt} e^{-\Ga_qt/2} \cos(\De M_q t/2)  ~, \nonumber\\
g_{-}(t) &=&  i\, e^{-iM_qt} e^{-\Ga_qt/2} \sin(\De M_q t/2) ~,
\eea
and $M_q=(M_H+M_L)/2$.

A fruitful approach used to study the effect of decoherence on the evolution of the $B$ system, as introduced in \cite{Bertlmann:1996at} for the $B_d$ 
mesons, is the modification of the interference term of the expression denoting the decay of the meson state $\ket{\Psi}$ into its final products $f_1$
and $f_2$, see Eq. (\ref{asl}), by a parameter $1-\zeta$. Here $\zeta$ comes from the phenomenological modelling of the process of decoherence (which originates from 
the effect of the environment on the evolution of the $B$ system) and takes possible values from 0 to 1, where 0 corresponds to complete coherence and 1 
to complete decoherence. This phenomenological prescription of the range of $\zeta$ comes from the understanding that decoherence is a form of noise 
inherent to the systems evolution. Thus $\zeta=0$ and 1 would correspond to zero and maximal noise, respectively.

From the probability of the decay of the meson state into its final products \cite{Carter:1980tk,Bigi:1981qs}, it can be shown that the ratio of the 
like-sign to opposite-sign dilepton events in $B_q\overline{B}_q$ decay is \cite{Bertlmann:1996at}
\begin{equation}
R_q \equiv \frac{N_{++} + N_{--}}{N_{+-} + N_{-+}}=\frac{1}{2} \l( \l| \frac{p}{q} \r|^2 + \l| \frac{q}{p} \r|^2  \r)
\frac{x_q^2 + y_q^2 + \zeta_q \l[ y_q^2 \frac{1+x_q^2}{1-y_q^2} + x_q^2 \frac{1-y_q^2}{1+x_q^2} \r] } 
{2+x_q^2-y_q^2 + \zeta_q \l[ y_q^2 \frac{1+x_q^2}{1-y_q^2} - x_q^2 \frac{1-y_q^2}{1+x_q^2} \r]},
\label{rq}
\end{equation}
where
\begin{equation}
x_q = \frac{\De M_q }{\Ga_q} \quad \mbox{and} \quad 
y_q = \frac{\De \Ga_q}{2 \Ga_q}.
\end{equation}
The parameter $R_q$ can be written as
\begin{equation}
R_q = \frac{\chi_q}{1-\chi_q}\,,
\end{equation}
where
\begin{equation}
\chi_q = \frac{x^2_q + y^2_q}{2(1+x^2_q)}\,.
\end{equation}

The semileptonic dilepton charge asymmetry is
\begin{equation} 
A^q_{sl} \equiv \frac{N \l(\bar{B}_q (t) \rightarrow l^{+} \nu_l X \r) - N \l(B_q (t) \rightarrow l^{-} \bar{\nu}_l X \r)} {N \l(\bar{B}_q (t) \rightarrow l^{+} \nu_l X \r) + N \l(B_q (t) \rightarrow l^{-} \bar{\nu} _l X \r)} =
\frac{|\frac{p}{q}|^2 - |\frac{q}{p}|^2}{|\frac{p}{q}|^2 + |\frac{q}{p}|^2}.
\label{asl}
\end{equation}

Using Eqs. (\ref{rq}) and (\ref{asl}), we have
\begin{equation}
\zeta_q = \frac{R_q \l(2+x^2_q-y^2_q \r)-\alpha_q \l(x^2_q+y^2_q \r)}{\l(\alpha_q - R_q \r)\l( y^2_q\frac{1+x^2_q}{1-y^2_q}\r) + 
\l(\alpha_q + R_q \r)\l(\frac{x^2_q\l(1-y^2_q\r)}{1+x^2_q}\r)}\,,
\label{zetaq}
\end{equation}
where
\begin{equation}
\alpha_q = \frac{1}{\sqrt{1-(A^q_{sl})^2}}\,.
\end{equation}

The effect of decoherence on the $B_d$ system evolution was studied in Ref. \cite{Bertlmann:1996at} in the flavour basis and in Ref. \cite{Dass:1997dd} in 
the mass basis. In Ref. \cite{Bertlmann:1997ei}, it was shown that the effect of decoherence is expected to be stronger in the flavor basis. In our 
nalysis we have made use of the flavour basis because if it turns out that the effect of decoherence in this basis is negligible then that understanding 
tends to the other basis as well.

\section{Results}
\label{sec:results}

In this section we present our results for the decoherence parameter $\zeta_q$, Eq. (\ref{zetaq}), for $B_d$ and $B_s$ mesons. 
The input parameters are given in Table \ref{input}.

\begin{table}
\begin{center}
\begin{tabular}{|c|c|}
\hline
\hline
$x_d = 0.770\pm 0.008$  & $x_s=26.49 \pm 0.29$ \\
$y_d = 0.0$  & $y_s=0.088 \pm 0.014$ \cite{lhcb-ys,Borissov:2013yha}  \\
$\chi_d = 0.1862 \pm 0.0023$  & $\chi_s= 0.499292 \pm 0.000016$  \\
\hline
\hline
\end{tabular}
\caption{Inputs that we use in order to obtain the decoherence parameter $\zeta_q$. When not explicitly stated, we take the inputs
from Particle Data Group \cite{pdg}.}
\label{input}
\end{center}
\end{table}

\subsection{Decoherence parameter $\zeta_d$ for $B_d$ system}

\begin{table}
\begin{tabular}{ccc}
\hline
\hline
$A^d_{sl}$  & $\zeta_d$\\
\hline
$0.0002 \pm 0.0031$ & $0.001 \pm 0.029$ \\
$0.0023 \pm 0.0026$ & $ 0.001 \pm 0.029$ \\
$0.0016\pm 0.0023$& $ 0.001 \pm 0.029 $ \\
\hline
\hline
\end{tabular}
\caption{Decoherence parameter $\zeta_d$ for $B_d$ meson system for different values of semileptonic dilepton charge asymmetry $A^d_{sl}$.
\label{resd}}
\end{table}

The values of $\zeta_d$ for various experimental values of semileptonic dilepton charge asymmetry $A^d_{sl}$ are presented in Table \ref{resd}.
The value of $A^d_{sl}$ in the first row of Table \ref{resd} is an average of all measurements performed at $B$ factories \cite{hfag1}. Adding the D\O\
measurement obtained with reconstructed $B_d$ decays \cite{Abazov:2012hha} to $A^d_{sl}$ given in the first row, yields the value given in second row. The 
latest dimuon D\O\ analysis separates the $B_d$ and $B_s$ contributions by exploiting their dependence on the muon impact parameter cut \cite{Abazov:2011yk}. The value of $A^d_{sl}$ given in third row is obtained by combining this latest result obtained by D\O\ with the  average $A^d_{sl}$  given in the second row \cite{hfag1}.  

It is obvious from  Table \ref{resd} that for all values of $A^d_{sl}$, $\zeta_d$ is 34$\sigma$ away from total decoherence. This is an improvement over 
the estimate made in \cite{Bertlmann:1996at,Bertlmann:2001iw}. Thus we see that the $B_d$ meson system is free from decoherence.

\subsection{Decoherence parameter $\zeta_s$ for $B_s$ system}

Here we present our results, pertaining to the effect of decoherence on $B_s$ systems for the first time. The values of $\zeta_s$ for various experimental 
values of semileptonic dilepton charge asymmetry $A^s_{sl}$ are presented in Table \ref{ress}.

The values of $A^s_{sl}$ in the first  and second row of Table \ref{ress} are obtained by D\O\ \cite{Abazov:2012zz} and LHCb \cite{lhcb-asl} collaborations, 
respectively, by measuring the time-integrated charge asymmetry of untagged $B_s \to D_s\, \mu \, X$ decays. The value of $A^s_{sl}$ obtained by LHCb is 
the most precise value to date. The value of $A^s_{sl}$ in the third row is obtained from CDF, D\O\ and LHCb analysis \cite{hfag1}. 

It is obvious from  Table \ref{ress} that depending upon the value of $A^s_{sl}$, $\zeta_s$ is 24$\sigma$ to 31$\sigma$ away from total decoherence. For the most 
precise data, obtained from LHCb \cite{lhcb-asl}, the deviation from total decoherence is seen to be 31$\sigma$. Thus we see that like the $B_d$ meson 
system, $B_s$ is also free from decoherence.

\begin{table}
\begin{tabular}{ccc}
\hline
\hline
$A^s_{sl} $  & $\zeta_s$\\
\hline
$-0.0108 \pm 0.0072 \pm 0.0017$ \cite{Abazov:2012zz}& $-0.023 \pm 0.042$\\
$-0.0024 \pm 0.0054 \pm 0.0033$ \cite{lhcb-asl}& $-0.004 \pm 0.032$\\
$ -0.0119 \pm 0.0038$ \cite{hfag1} & $ -0.028\pm0.036$ \\
\hline
\hline
\end{tabular}
\caption{Decoherence parameter $\zeta_s$ for $B_s$ meson system for different values of semileptonic dilepton charge asymmetry $A^s_{sl} $.
\label{ress}}
\end{table}

\section{Conclusion}
\label{sec:concl}
We study the impact of decoherence on $B$ meson systems with specific emphasis on $B_s$. For consistency we also study the $B_d$ mesons based on the most 
recent data. We find, using the currently available data, that the $B_d$ mesons are $34\sigma$ away from total decoherence, while the  $B_s$ mesons are $24\sigma$ to $31\sigma$ away from total decoherence,
depending upon the input data on the semileptonic dilepton charge asymmetry.
Using the most recent experimental data, our results thus prove that neutral mesons are free from decoherence effects. Therefore, this provides a very
useful laboratory for testing the foundations of quantum mechanics. 



\end{document}